\documentclass[sigconf, authorversion=true]{acmart}
\usepackage[english]{babel}
\usepackage{booktabs} 
\usepackage{multirow}
\usepackage{parskip}
\usepackage{balance}
\usepackage{array}
\usepackage[para,online]{threeparttable}

\copyrightyear{2018}
\acmYear{2018}
\setcopyright{acmcopyright}
\acmConference[SAICSIT '18]{2018 Annual Conference of the South African Institute of Computer Scientists and Information Technologists}{September 26--28, 2018}{Port Elizabeth, South Africa}
\acmBooktitle{2018 Annual Conference of the South African Institute of Computer Scientists and Information Technologists (SAICSIT '18), September 26--28, 2018, Port Elizabeth, South Africa}
\acmPrice{15.00}
\acmDOI{10.1145/3278681.3278692}
\acmISBN{978-1-4503-6647-2/18/09}

\newcommand\MyBox[2]{
  \fbox{\lower0.75cm
    \vbox to 1.7cm{\vfil
      \hbox to 1.7cm{\hfil\parbox{1.4cm}{#1\\#2}\hfil}
      \vfil}%
  }%
}

\begin{document}
\title{A Network Topology Approach to Bot Classification}

\author{Laurenz A. Cornelissen}
\orcid{0000-0001-7864-3143}
\affiliation{%
  \institution{Computational Social Science Group\\
  Centre for AI Research\\
  Department of Information Science\\
  Stellenbosch University}
  }
\email{alducornelissen@sun.ac.za}

\author{Richard J Barnett}
\orcid{0000-0003-4133-4587}
\affiliation{%
  \institution{Computational Social Science Group\\
  Centre for AI Research\\
  Department of Information Science\\
  Stellenbosch University}
  }
\email{barnettrj@acm.org}

\author{Petrus Schoonwinkel}
\orcid{0000-0002-2148-8880}
\affiliation{%
  \institution{Computational Social Science Group\\
  Department of Information Science\\
  Stellenbosch University}
  }
\email{petrus.schoonwinkel@gmail.com}

\author{Brent D. Eichstadt}
\orcid{0000-0001-5080-1947}
\affiliation{%
  \institution{Computational Social Science Group\\
  Department of Information Science\\
  Stellenbosch University}
  }

\author{Hluma B. Magodla}
\orcid{0000-0002-5415-705X}
\affiliation{%
  \institution{Computational Social Science Group\\
  Department of Information Science\\
  Stellenbosch University}
  }

 \renewcommand{\shortauthors}{L.A. Cornelissen \emph{et al.}}

\begin{abstract}
Automated social agents, or \emph{bots} are increasingly becoming a problem on social media platforms. There is a growing body of literature and multiple tools to aid in the detection of such agents on online social networking platforms. We propose that the social network topology of a user would be sufficient to determine whether the user is a automated agent or a human. To test this, we use a publicly available dataset containing users on Twitter labelled as either automated social agent or human. Using an unsupervised machine learning approach, we obtain a detection accuracy rate of 70\%.
\end{abstract}

%
%

\ccsdesc[500]{Security and privacy~Social network security and privacy}
\ccsdesc[300]{Human-centered computing~Collaborative and social computing}

\keywords{Automated Social Agent Detection, Social Network Theory, Unsupervised Machine Learning, Twitter}

\maketitle
\section{Introduction}
\label{S:Introduction}
The use of automated agents, or \textit{bots}, are increasingly prevalent on social media platforms \citep{Ferrara2016}. There are many examples of harmless, and even helpful bots, such as a massive Star Wars obsessed bot-net \citep{Echeverria2017}, or countless client relations management bots, deployed by corporations to help deal with clients on online social networks (OSNs).

To create bots on OSNs is a relatively basic task and has been used for many years. Bots were employed mostly with harmless use-cases, but the applications have since spread to potentially more damaging activities. For a fee, organisations can artificially boost their profile, products, or ideas on OSNs by utilising such bots to tweet, post, favourite, retweet, follow, comment, befriend or reply. Similarly, a person can make themselves seem more trustworthy, credible or noteworthy, and ideas can also be boosted or quashed artificially. These objectives play well into the political arena, where the popularity and credibility of personalities and ideas are of central concern. \citet{Cresci2017} argue that there is evidence of a new paradigm in bot development, which they term `social bots'. This new wave of bots are smarter and have more advanced objectives. They attempt to evade detection by increasingly emulating expected social behaviour, by mimicking human behaviour. These social bots are much harder to detect, or at least, to distinguish from humans.

\citet{Echeverria2017} offer a brief overview of such `threatening' bot activities on Twitter: spamming, fake trending topics, opinion manipulation, astroturfing\footnote{Astroturfing is the act of sponsoring a campaign to look like a legitimate grass-roots movement. Astroturf is an artificial type of grass, thus the name.}, fake followers and API contamination. The most notorious example in recent memory of such activities was with the British referendum on European Union membership,\footnote{Popularly referred to as Brexit, which is an amalgamate of British Exit.} where both sides of the debate included bot activities \citep{Bastos2017}. Another is the 2016 US Presidential election \citep{Shao2018}. This is an interesting turn from social media's earlier hopeful and optimistic coming of age with the Arab Spring \citep{Howard2011, Gerbaudo2012}. Indeed, recent warnings from historians are highlighting the argument that connectedness does not necessarily lead to togetherness \citep{Ferguson2017}.

The above are examples of political interference in relatively homogeneous and politically stable developed nations. One can argue that these elections are extremely consequential worldwide, and therefore attract attention from various players, who are willing and able to interfere. Thus, elections in a developing nation may not attract such attention and levels of artificial interference. This is however not a good assumption, since there is evidence of interference by both local and international players in the South African political landscape on social media \citep{Findlay2018}.

Since 2016, South Africa has been experiencing its first large scale political interference on OSN platforms. There are multiple accounts of astroturfing involving pressure groups, and social bots driving political messages for a controversial politically connected business family, and `fake news' spreading about various topics. Many hundreds of `bot' accounts and misinformation networks have been uncovered by journalists who join a growing community of researchers attempting to solve the difficult, and increasingly important, issue of bot detection on OSNs \citep{Fraser2017}.

The scale of the worldwide issue has even attracted the attention of DARPA who hosted a competition for teams to detect automated social agents \citep{Subrahmanian2016}. The outcomes of the competition revealed that bot detection should be a semi-supervised process, a combination of crowd-sourcing and feature-based detection, which are discussed later in this paper.

The rest of the paper will firstly review the relevant literature on detecting bots on OSNs. From the literature, we find two major approaches to bot detection that can be classified as feature based and network based. We then propose social network topology as viable feature vector in distinguishing between bots and humans. We then propose an unsupervised learning methodology using social network topology. Finally, in the last section, we report the results of the classification.

\section{Bot Detection}
An exhaustive overview of bot detection methods is beyond the scope of this study, and it is sufficiently covered by \citet{Ferrara2016}, who offer a helpful taxonomy of detection methods. \citet{Gilani2017} pointed out that many of these methods have claims of highly successful classification attempts, but few offer their methods and/or datasets with which to benchmark. The next sections offer a brief and general overview of previous approaches to bot classification.

\subsection{Crowd Sourcing}
The first is a non-computational solution, which relies on humans to detect bots on social platforms. The idea is that a large group of people would act like a mass Turing test to classify users as human or bot. There is a high success rate and near zero false positives, since humans are able to pick up on social nuances that are difficult to encode \citep{Wang2012}.

There are, however, multiple drawbacks to this approach, most notably the cost to scale implications. Only large social networking platforms are able to afford a group of expert analysts. Moreover, there are clear issues with privacy concerns when user profiles and data are exposed to people for annotation purposes. More modern bots are able to easily appear as more human, so annotation becomes inceasingly difficult. Especially if no supplementary automated methods are used.

Instead of relying on humans to detect anomalous features of OSN profiles, the problem would be well suited to a machine learning approach. The next sections explores such approaches.

\subsection{Feature-based Detection}
\label{S:Feature-Based}
Feature-based detection methods distil data from OSN users into analysable features. Given sufficient examples of bots and humans, multiple features can be recorded from each category to identify significant differences between them. These features mostly rely on behavioural data of the agents for classification, however, it is reasonable to combine multiple features.

This problem lends itself to the implementation of machine learning libraries in order to help classify using multiple features. Bot\-ometer, the first publicly available classifier, uses a random forest ensemble supervised learning method \citep{Davis2016}.\footnote{The name was originally \textit{BotOrNot}, but was since changed to \textit{Botometer}, see \url{https://botometer.iuni.iu.edu}.} The classifier uses multiple features from the user's network, user profile, friends, temporal features, content and sentiment. The exact implementation is not publicly available, but \citet{Gilani2017} attempted a reproduction and extension to the work of \citet{Davis2016} and made the methodology and data open access.

This type of bot detection and classification work enjoys the most success. There are some notable new developments since the review of \citet{Ferrara2016}. Inspired by digital DNA sequencing, \citet{Cresci2017} developed `Social Fingerprinting' as a way to classify bots and human agents.
DNA sequencing creates a set of similarity curves to which honest and dishonest nodes adhere. The approach has a 92.9\% accuracy \citep{Cresci2017}.

Since the platforms are fundamentally social, many researchers have developed methods that use these social characteristics in bot detection. The next sections explore these characteristics in the form of graph-based methods.

\subsection{Graph-based Detection}
The intuition with graph-based approaches is that because the platform is social, the interaction between users of the platform generate trace data of their interactions. This trace data can be used to uncover the social aspect of the user behaviour, which can be used to differentiate between bots and humans. Generally the trace data is envisioned as a network graph.

There are two approaches to generating graph data. The first uses the content interaction between users of a social media platform to create a graph. The second approach uses the relational data, such as following and friending, to create a network. Each approach is discussed in more detail below.

\subsubsection{Content-based}
Content-based graphs are used as feature variables as described in Section~\ref{S:Feature-Based}. Examples include \citet{Varol2017} who use content networks produced by retweet, mention and hash-tag co-occurrence of users to produce network structures \citep[also see][]{Davis2016, Gilani2017, Cresci2017}. For each network, various network features are calculated, including indegree, outdegree, density and clustering coefficient. These measures are then combined with other features including user meta data, friends data, content, sentiment and timing. A recent unique content based method is by \citet{Shao2018}, who used content interaction to investigate the dynamics of the spread of false information. They observed, through k-core decomposition, that social bots proliferate at the fringes of the interaction network.

Most research where the graph is used as the key, or only, feature is  categorised as relation-based detection. The next section reviews this research agenda.

\subsubsection{Relation-based}
Research using relation-based graphs for bot detection is dominated by what is labelled Sybil research.\footnote{The original usage of the term `Sybil' is by \citet{Douceur2002}.} In Sybil research the key assumption is that trust is more difficult or expensive to establish between a bot and a human. This means that the number of connections between humans and bots would be less than within each group. In other words, people will not readily follow bots. This assumption lends itself to community detection methods, which is well developed within graph theory. The intuition is that if one can identify nodes belonging to the same community as a confirmed account, either bot or human, they should inherit the label.

These methods first identify a confirmed honest or Sybil seed node, then a surrounding relational network structure is elicited. From this network structure, they use various community detection algorithms to classify all nodes in the network into a community \citep{Viswanath2010}. Those nodes classified into a community with the seed node inherit the label; honest or Sybil. Examples of such methods include SybilGuard \citep{Yu2008}, SybilLimit \citep{Yu2008a}, SybilInfer \citep{Danezis2009}, and more recently SybilFuse \citep{Gao2018}.

Bots are able to establish connections to humans either through effective disguise or poor due-diligence by the human agents. Many researchers found that the trust assumption is not as valid as originally anticipated.\footnote{Although the use of trust is a broad application of actual human trust. A better proxy for this might be a type of threshold.} \citet{Yang2012} investigated the effectiveness of \textit{criminal supporter} accounts which aim to infiltrate and hide actual bot accounts, offering proxy attack edges.\footnote{Attack edges are connections made from a sybil account to an honest user, through which attacks can be performed.} \citet{Edwards2014} experimentally tested whether a known bot would be perceived differently from a human in terms of communication factors. They found no difference between communication quality of bots and human agents, except for a measure of attraction. \citet{Boshmaf2011} tested the level of social penetration by building their own bot-net, which achieved as high as 80\% penetration. \citet{Ghosh2012} highlighted the effect of the \textit{social capitalists}---human agents who always follow back---who make it easy for bot-nets to penetrate social networks and increase attack edges between bot communities and humans. \citet{Bilge2009} investigated how easy it would be to automate cloning attacks. Both their simplistic and more advanced methods offered positive results. The assumption of strong trust networks is, therefore, difficult to rely on. \citet{Aiello2014} conducted a social experiment where they determined how a bot can gain influence on social media.

Bots are, therefore, becoming more socially complex and thus more difficult to distinguish from honest users of the platform. The assumptions of previous methods are increasingly becoming ineffective in detecting such bots \citep{Cresci2017}. The most successful methods are those that combine multiple methods into a comprehensive detection and classification regime.

Consequently, the options for future research are two-fold. Firstly, to determine effective combinations of the above explored methodologies, such as \citet{Varol2017, Cresci2017, Gilani2017}, and secondly, to develop new specialised methodologies and data gathering techniques, such as new measures of on-line agents. Here we propose a new specialised methodology, which could be included in future blended methodologies. The proposed method is an unsupervised machine learning approach, using existing clustering methods on features extracted from a K-2 network of a target node.\footnote{This is defined in section 3.2, but a K-2 network is a network that spans two steps from an original seed node (ego).}

\subsubsection{Issues with prior approaches}
Existing classification methods have two key issues. Firstly, many are sensitive to the training datasets used to validate the method---especially methods relying on supervised machine learning. Secondly, since research about bot detection is published---an obvious reason why Twitter does not do so---the detection methods are available to those who wish to avoid detection.

By using human annotators, \citet{Gilani2017} determined that Bot\-ometer could only achieve accuracy measure of 48\% when presented with a new dataset. This is compared to a claimed 95\% accuracy rate with previous training data by by \citet{Davis2016}, which was later reduced to 88\% by \citet{Varol2017}.

Supervised machine learning methods are sensitive to biases in training datasets. For instance, the Botometer classifier is based on an ensemble learning random forest algorithm. Although they do check for over-fitting, it is difficult to completely alleviate the sensitivity, and automated social agents are constantly evolving, which necessitates new features to be developed to cover for such evolution. A less covered and less substantiated method is the \textit{botcheck.me} classifier,\footnote{See https://botcheck.me/} which offers an example of specialisation classification. They use machine learning to classify automated agents, but are explicit about the narrow application, which is politically orientated bots. The botcheck.me classifier is specifically trained on `politically orientated' bots, instead of attempting a generalist approach such as BotOrNot. The prominence of more complex approaches, for example SybilFuse by \citet{Gao2018}, is a response on the oversimplification of existing methods in bot detection.

\citet{Cresci2017} determined that bots evolve as new detection methods arrive. It is reasonable to expect that bot detection methodologies that are well publicised as being effective would enable those avoiding detection to learn how to avoid it \citep{Varol2017}. The most well known publicly available method of bot detection is Botometer, since generating coverage in news articles and research `fact tanks' such as Pew Research Centre \citep{Gramlich2018}. If this is the reason for the low correlation between human annotators and the Botometer detection method's accuracy, then it deteriorated at a surprisingly rapid pace. The Botometer paper by \citet{Davis2016} was published in April of 2016, and a year and a half later, \citet{Gilani2017} determined that the accuracy of the method dropped by an estimated 47\%.

\subsection{Network Topology Approach}
We propose here a new specialised approach, using ego-centred network topology as feature vectors for unsupervised machine learning.

The appeal of network topology as a feature is that it is able to capture complex social structure, which social bots would struggle to effectively emulate. A large body of work exists, which investigates the complexities of social networks, like small-world and scale-free properties \citep{Barabasi2002,Newman2003}. At risk of oversimplifying the field, people do not form social relations in a random manner. The establishment of connections have local biases, yet the macro effect of these biasses have complex outcomes. For instance, the preferential attachment model is put forth as the reason why networks have scale-free properties \citep{Barabasi2002}. Preferential attachment implies that nodes with more relations are more probable to receive more relations. This is also called the Matthew effect, or the \textit{rich get richer} effect.

If we imagine a simplistic bot befriending people at random, the degree distribution would most probably follow a normal distribution, whereas real social networks almost exclusively follow a power law distribution \citep{Barabasi2002}. This is easily remedied by the designer of the social bot by adding a parameter to prefer befriending people with more followers and friends. The network topology, therefore, seems to be easily circumvented by a well designed social bot. However, these bots can not control the interactions by the bots alters. For instance, if they are only ever followed back by people who automatically follow back all requests, such as how \citet{Ghosh2012} reports, then this would show up clearly in the topology as an extraordinarily dense and large network. Emulating such complexities of social networks become increasingly more difficult when considering the many idiosyncratic social behaviours of users.

The next section expands on the methodology, specifically the network topology methods that were used in developing the feature vector. We also highlight the classification methodology used.

\section{Methodology}
To test the accuracy of the proposed method, an annotated dataset containing both bots and humans is required. Given a list of identifiers, a two step crawler, which we label the K-2 crawler, is utilised. Multiple network measures are then calculated for each graph as a feature of the seed node. These measures are then used in a clustering procedure. The inherited labels from the clusters are then compared to the original annotations to determine the accuracy of the method. The following sections describe this methodology in more detail.

\subsection{Dataset}
Since we need an annotated dataset containing only a Twitter user identifier, and its annotation as bot or not, there are a few publicly available datasets that satisfy these criteria. The most accessible dataset for our purposes was made available by \citet{Varol2017}. The \citet{Varol2017} dataset contains Twitter user identifiers and their corresponding classification; 1 = Bot, 0 = Not. The labels were assigned as 1 when the user has a classification score of 0.5 and above \citep{Varol2017}. With the identifiers it is possible to crawl the follower networks. The original dataset contained 2573 identifiers containing 68\% (1747) nots and 32\% (826) bots. After checking which of the accounts are still active, the revised dataset contains 2178 identifiers, which is a 15\% (395) reduction. The split between the annotations are similar with 67\% (1462) annotated as Nots and 33\% (671) as Bots.

\subsection{K-2 Crawler}
\begin{figure}
	\includegraphics[width=0.45\textwidth]{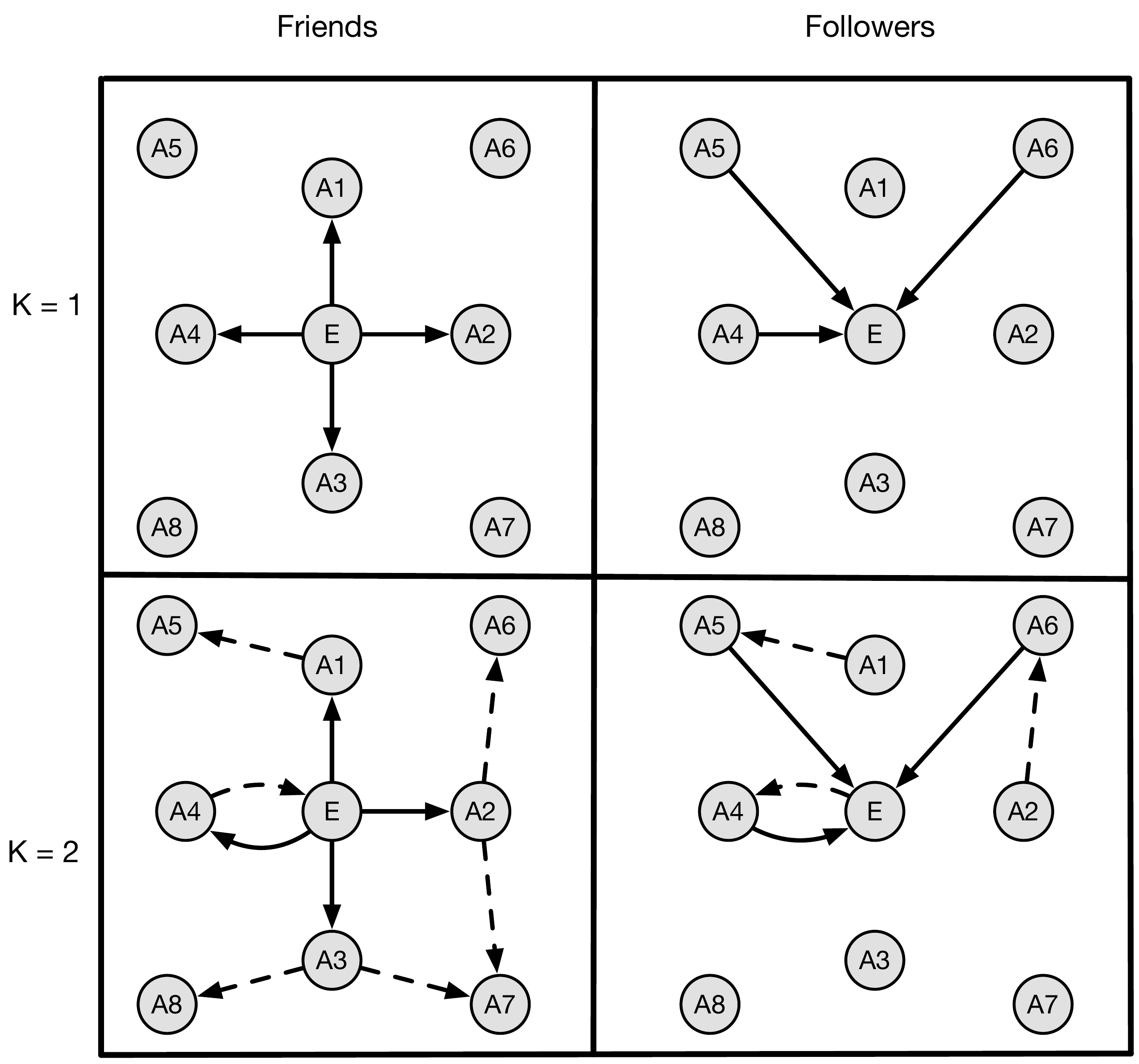}
	\caption{Descriptive Diagram of Crawl Variants}
	\label{F:K2}
\end{figure}

The objective of the crawler is to be able to gather the surrounding relational network structure of the ego. Previous studies have also generated such relational graphs around a focal node, but we propose a wider crawl. Most prior research stop the crawl at K=1 from ego, and some include both friends and followers. To illustrate the variants of the crawl we refer to Figure~\ref{F:K2}. A simplistic crawl can vary in two parameters, direction (friends or followers) and k (steps). As an example, \citet{Yang2013} did both the top two quadrants by crawling K=1 on both the friends and followers of each user. Many researchers take the count of followers and friends in order to calculate the ratio. Actually retrieving the user identifiers of the friends or followers make it possible to determine, as in the case of \citet{Yang2013}, different measures such as reciprocity. Extending the crawl to K=2, makes even more measurements possible. In practice a K=3 crawl is impractical on Twitter, since this could theoretically crawl the whole network with a median distance of 4.12 reached in 2010 already \citep[p.~594]{Kwak2010}. We, therefore, perform a crawl as illustrated in the bottom-left quadrant in Figure~\ref{F:K2}, which makes more nuanced graph measures possible. We are only interested in friends of egos and alters, and not followers, since we are interested in the `choices' of the users, and the reaction to those choices. For instance, we are interested to know whether, if an ego follows an alter, does the alter follow back, and among those that follow back, do they follow each other.

\subsection{K-1 Graph}
We also reduced the original graph to an ego-centric, or K-1, graph. This graph utilises the full graph's K-2 crawl and then reduces the number of nodes in the dataset by doing a k-core decomposition. The goal is to gain insight about the structure, function, and composition of network ties around the ego specifically \citep{Wasserman1994}. Since \citet{Boshmaf2011} showed that it is possible for a bot to penetrate true social structures, we are cognisant of the possibility that the actual social topology might distort the structure of a possible bot. Reducing the K-2 to a K-1 network will reduce the added noise of the alters' full networks and concentrate the measurements on the seed node, thus called an ego-centric network, since it centres around the ego node.

Since social networks can not be compared directly, we have to reduce them to certain features to compare them with. The following section outlines the measures of interest here.

\subsection{Graph Measures}
While there are many relevant graph measures,\footnote{ For a good start see \citet{Wasserman1994}.} we are more interested in a few general and easily interpretable measures as a proof of concept in using graph structure as a means to classify users on Twitter. Table~\ref{T:Network_Measures} offers a brief overview of the measures used. Since the K-2 networks can become excessively large, we can not use computationally expensive measures. For convenience the measures are summarised in Table \ref{T:Network_Measures}.

\begin{table*}[]
	\centering
	\footnotesize
	\caption{Network Measures}
	\label{T:Network_Measures}
	\begin{tabular}{@{}p{0.2\textwidth}p{0.65\textwidth}p{0.15\textwidth}@{}}
		\toprule
		\textbf{Measure}				& \textbf{Description} 							& \textbf{Source}\\ \midrule
		Size 							& The number of nodes in the graph. & \citep[p.5-8]{Harris2008}\\
		Global Clustering Coefficient  & ``The fraction of paths of length two in the network that are closed''. i.e. Whether a friend of a friend is a friend. & \citep[p.~199]{Newman2010}    \\
		Local Clustering Coefficient  & The same as the global measure, but measured for a focal node. & \citep[p.~201]{Newman2010} \\
		Indegree Graph Centralization & A graph-level measure of the number of edges directed towards the nodes in a graph.& \citep[p.~175-177]{Wasserman1994}\\
		Outdegree Graph Centralization & A graph-level measure of the number of edges directed from the nodes in a graph.& \citep[p.~175-177]{Wasserman1994} \\
		Degree Graph Centralization   & A graph-level measure of the number of edges directed to and from the nodes in a graph.& \citep[p.~175-177]{Wasserman1994}\\
		Indegree Graph Centrality   & The number of edges that are directed towards a single node.& \citep[p.~178,199-202]{Wasserman1994}\\
		Outdegree Graph Centrality   & The number of edges that are directed from a single node to other nodes in the graph.& \citep[p.~178,199-202]{Wasserman1994}\\
		Degree Graph Centrality     & The number of edges directed from and at a single node in a graph.& \citep[p.~178,199-202]{Wasserman1994}\\
		Density             & Ratio of the amount of edges and the amount of possible edges in the graph.& \citep[p.~165]{Wasserman1994}\\
		Reciprocation          & The proportion of mutual connections in a directed graph.& \citep[p.~515]{Wasserman1994}\\
		Assortativity          & Also known as assortative mixing. Assortativity is the preference for a graph's nodes to attach itself to other nodes that are similar to it. The similarity, in this case is measured by degree. & \citep{Newman2003}\\
		Articulation Points       & Nodes that if removed, increases the number of connected components in a graph. Also known as cut vertices.& \citep{Italiano2012}                                                                      \\ \bottomrule
	\end{tabular}
\end{table*}

\subsubsection{Network Size}
Network size is a measure of the number of nodes (users) in the network. Network size is not a particularly useful measure of network topology, but it is an important measure for normalisation of the data.

\subsubsection{Density}
Density is the measure of how well connected a network is. It is expressed as a proportion of relations out of the theoretical maximum possible relations in the network. It is also a control measure of other topology measures such as centrality, since the measures are sensitive to the density and the size of the network.

\subsubsection{Clustering Coefficient}
Clustering coefficient is the simple, yet informative, measure of how transitive a relation is between any three nodes. It, therefore, measures whether a friend of a friend is also a friend. Social networks tend to have a higher clustering coefficient than a random network of the same size \citep[p.~50]{Barabasi2002}. We calculate both the global clustering coefficient and the local clustering coefficient, to capture the transitivity of the network as a whole as well as the ego self.

\subsubsection{Centrality}
Centrality is a useful measure of importance of nodes in a graph. There are many measures of centrality, but the simplest measure is to count the number of connections to a node. These connections could be either undirected, or directed in or out of a node. A node has the highest degree centrality when it has more relations to it than any other node. Some nodes might have high indegree but low outdegree. For instance, celebrities on Twitter tend to have more people following them than they themselves follow. Therefore, celebrities might have high indegree but low outdegree. Simply using degree, without direction, would not offer this type of nuance. A simplistic bot on the other hand might have very high outdegree but very low indegree, since most people will not follow a bot back. A social capitalist on Twitter could be identified by a balanced indegree and outdegree, since they reciprocate every relation.

\subsubsection{Graph centralisation}
Graph centralisation is a measure of how much the degree count is concentrated in a single node. Centralisation can be calculated by either taking the degree of nodes (bot incoming and outgoing relations), or it can use the direction of the relation. A graph is perfectly centralised if a single node contains all the relations. A graph is completely decentralised if all the nodes have an equal degree.

\subsubsection{Reciprocity}
Reciprocity is measure of how many relations in the network are reciprocated. It is anticipated that bots would have lower reciprocity than human users.

\subsubsection{Assortativity}
Assortativity is a measure of the tendency of nodes to connect to similar nodes, i.e. homophily. Usually assortativity is calculated using a node variable such as gender or age, we use degree as a measure of similarity. We therefore test the tendency of high degree nodes to only follow other high degree nodes, and inversely low degree nodes to only follow fellow low degree nodes. This can be likened to celebrities only following celebrities.

\subsubsection{Articulation Points}
Articulation points are the points which, if removed, increase the number of connected components in a network. Social networks tend to have higher redundancy, which translates to fewer articulation points than random graphs.

Each seed node in the Varol dataset was crawled using the K-2 crawler. The resulting network of each crawl was copied and reduced to an K-1 network. We, therefore, have two networks representing each identifier from the dataset. The metrics were then calculated for each of these networks. There are, therefore, two sets of network measures for each identifier. These measures were then used as feature vectors in the clustering methodology.

\subsection{Clustering Methodology}
The clustering methodology involves various steps. Firstly, a visual assessment of clustering tendency (VAT) is performed as prescribed by \citet{Bezdek2002}. This offers a means to determine the most applicable distance measures. Secondly, four distinct clustering methods are considered in order to determine which method offers the best performance in classifying users.

\subsubsection{Distance Measures}
There are multiple distance measures to consider. In addition to standard euclidean distance, three generally applicable measures are, Pearson, Spearman, and Kendall. To determine which distance measures perform best, we inspect the performance of the measures in producing clear clusters. To do this, we follow the VAT procedure from \citet{Bezdek2002}. Firstly, we compute the dissimilarity matrix between the objects based on the network topology measures. This dissimilarity matrix is then reordered to place similar objects closer to each other, producing an ordered dissimilarity index which is shown in Table~\ref{T:IDM}.

\begin{table*}[]
	\centering
	\caption{IDMs of the Various Distance Measures}
	\label{T:IDM}
	\resizebox{0.9\textwidth}{!}{%
		\begin{tabular}{@{}lcccc@{}}
			\toprule
			\multicolumn{1}{c}{\textbf{}} & \multicolumn{4}{c}{\textbf{Method}} \\ \cmidrule(l){2-5}
			\multicolumn{1}{c}{\begin{tabular}[c]{@{}c@{}}Graph\\ Type\end{tabular}} & Euclidean & Kendal & Spearman & Pearson \\ \midrule
			&  &  &  &  \\
			& \multirow{7}{*}{\includegraphics[width=3.8cm]{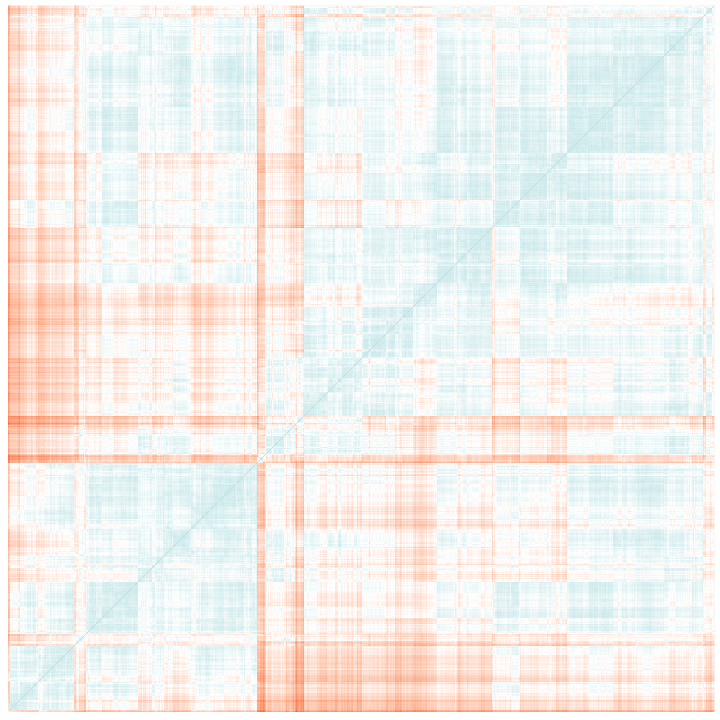}} & \multirow{7}{*}{\includegraphics[width=3.8cm]{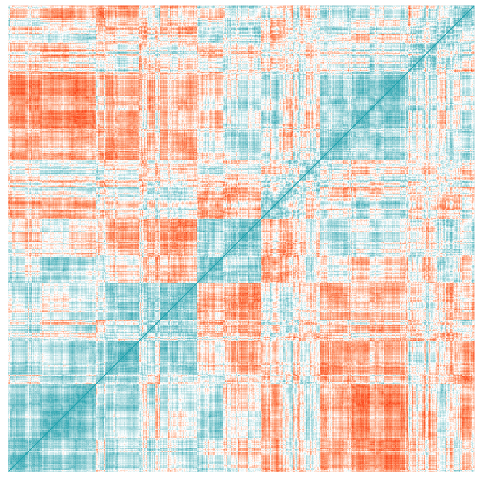}} & \multirow{7}{*}{\includegraphics[width=3.8cm]{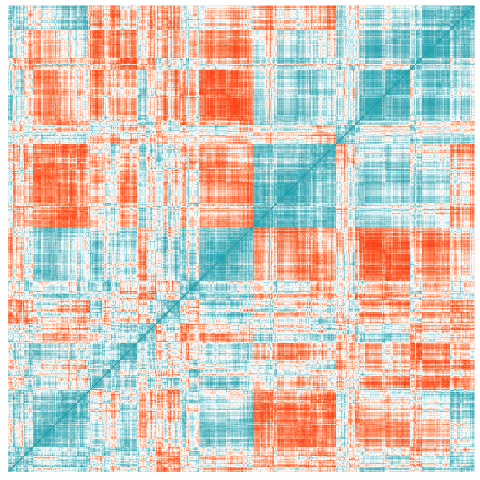}} & \multirow{7}{*}{\includegraphics[width=3.8cm]{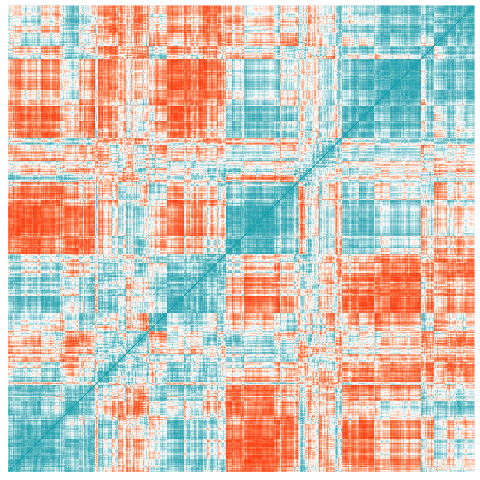}} \\
			&  &  &  &  \\
			&  &  &  &  \\
			&  &  &  &  \\
			&  &  &  &  \\
			Original (K-2) &  &  &  &  \\
			&  &  &  &  \\
			&  &  &  &  \\
			&  &  &  &  \\
			&  &  &  &  \\
			&  &  &  &  \\
			& \multirow{7}{*}{\includegraphics[width=3.8cm]{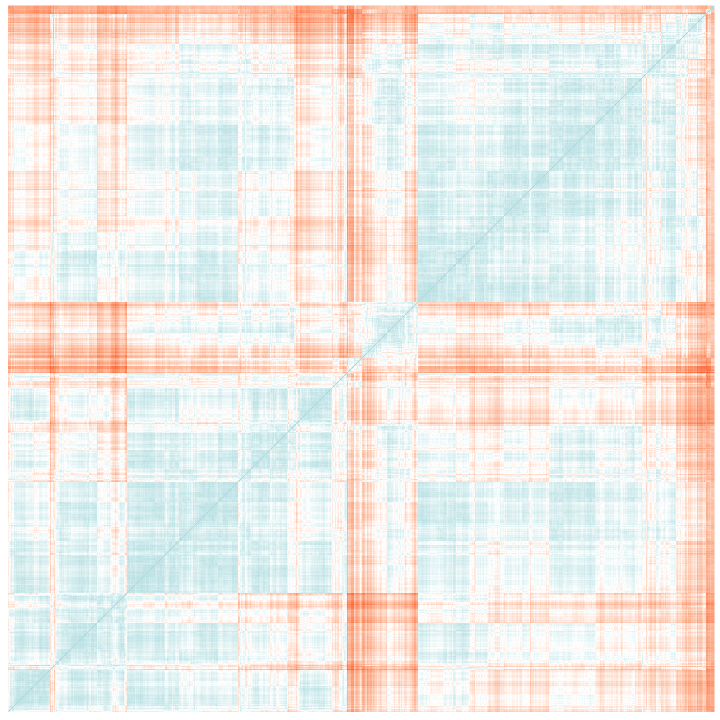}} & \multirow{7}{*}{\includegraphics[width=3.8cm]{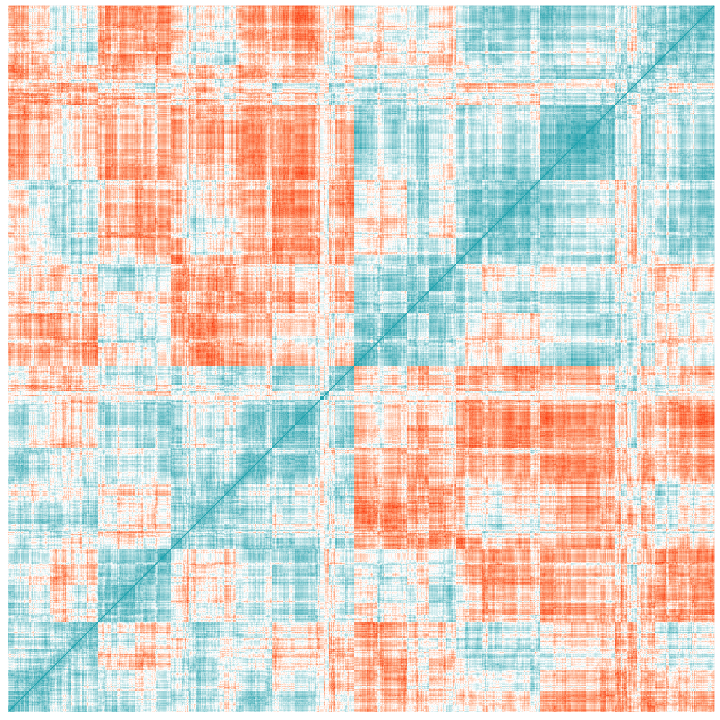}} & \multirow{7}{*}{\includegraphics[width=3.8cm]{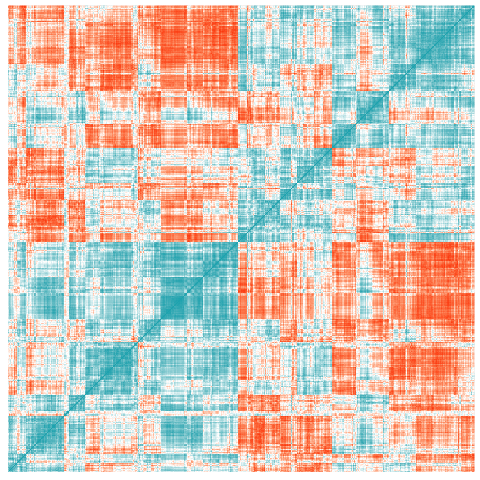}} & \multirow{7}{*}{\includegraphics[width=3.8cm]{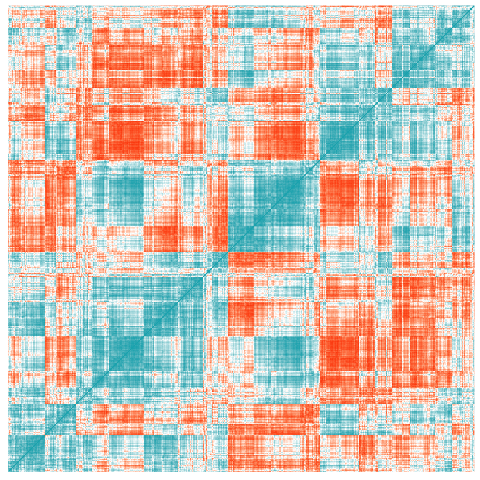}} \\
			&  &  &  &  \\
			&  &  &  &  \\
			&  &  &  &  \\
			&  &  &  &  \\
			K-1 (Ego-centric) &  &  &  &  \\
			&  &  &  &  \\
			&  &  &  &  \\
			&  &  &  &  \\
			&  &  &  &  \\
			&  &  &  &  \\
			\bottomrule
		\end{tabular}%
	}
\end{table*}

A visual inspection indicates that the two correlation measures offer the best definition of clusters---Pearson and Spearman.

Pearson measures the difference between the variables of each observation, while Spearman ranks the observations based on the differences between the variables. In other words, both correlations evaluate to the relationships between two nodes but differ in that Spearman correlation measures variables that change together but not necessarily at the same time.

Therefore, both Spearman and Pearson are used as options for distance measures in the clustering procedure.

\subsubsection{Clustering Methods}
To determine the most appropriate clustering method and number of clusters, we use cluster validation procedures as suggested by \citet[p~138-141]{Kassambara2017}. Both internal and stability validation methods were performed. Internal cluster validation procedures evaluate the results of clustering procedures on three measures: connectivity, Dunn and silhouette. Stability measures include: average proportion of non-overlap (APN), average distance (AD), average distance between means (ADM) and the figure of merit (FOM) \citet[p~152]{Kassambara2017}. All the measures were performed on a representative sample of the dataset, roughly 10\% of the full Varol dataset, since the measures require an exponential amount of processing power as the number of nodes increase. The results of these validation methods suggest that the most applicable clustering methods would be partitioning around medoids (PAM), fuzzy analysis clustering (FANNY) and agglomerative nesting (AGNES), with two clusters as the most optimal number. Each method is briefly explained below.

\paragraph{K-means (Partitioning Around Medoids)}
The first clustering method, PAM, is used to partition the data into K groups, where K is the amount requested by the analyst. This groups the nodes around the two or more largest mean distances between nodes. For the purpose of analysing this dataset, two points were chosen to cluster Bots and Nots into groups that could be easily distinguished \citep{Wagstaff2001}.

\paragraph{Fuzzy Analysis Clustering}
FANNY groups nodes based on the probability of belonging to a specific cluster. The algorithm calculates the distance between nodes and the centre of a cluster and assigns a coefficient to each node. The clusters were computed using the R statistical language according to the guide provided by \citep{Kassambara2017}.\footnote{All analysis was conducted using R.}

\paragraph{Agglomerative Nesting Method}
AGNES groups nodes based on their similarity coefficient. The algorithm merges nodes in series of steps, creating a larger cluster with each step until only one remains, leaving behind a dendrogram representation of the nodes \citep{Kassambara2017}.

The two distance measures (Pearson and Spearman), three clustering methods (FANNY, AGNES and PAM) and two graph structures (full and K-1), result in twelve distinct methods to use in classifying the observations. Before proceeding to the performance results, the methodology for measuring the performance of the 12 methods needs to be defined. The next section covers the measures of clustering performance.

\subsection{Clustering Performance Methodology}
A confusion table is a helpful framework to measure the performance of classification methods.

Table~\ref{T:CONFT} is an example of a \textit{Confusion Table}. To use the confusion table, there must be a predicted outcome and an actual outcome. Each clustering method offered two clusters, each observation was placed within a cluster. There is no a-priori way of knowing what the clusters consist of, nor is it known if it is actually differentiating between bots and humans. For each observation the original label from \citet{Varol2017} used as comparison. Recall that each observation was classified as either 1 or 0, indicating (respectively) bot or not. We therefore know the actual class of the observation, and can then compare the prediction of the clustering method against the original Varol labels.
\begin{center}
	\noindent
	\renewcommand\arraystretch{1.5}
	\setlength\tabcolsep{0pt}
	\label{T:CONFT}
	\begin{tabular}
		{c >{\bfseries}r @{\hspace{0.7em}}c @{\hspace{0.4em}}c @{\hspace{0.7em}}l}
		\multirow{10}{*}{\rotatebox{90}{\parbox{1.1cm}{\bfseries\centering Cluster}}} &
		& \multicolumn{2}{c}{\bfseries Actual Label} & \\
		& & \bfseries Bot & \bfseries Not & \bfseries Total \\
		& Bot$'$ & \MyBox{True}{Positive} & \MyBox{False}{Positive} & BOTS$'$ \\[2.4em]
		& Not$'$ & \MyBox{False}{Negative} & \MyBox{True}{Negative} & NOTS$'$ \\
		& Total & BOTS & NOTS &
	\end{tabular}
	\captionof{table}{Confusion Table}
\end{center}

Table~\ref{T:CONFT}, shows the four possible outcomes for a classifier. When both the clustering method and the actual labels indicate \textit{Bot}, the classification is a true positive (TP). If the classification indicates that an observation is classified as \textit{Not}, where the actual label is \textit{Bot}, the result of the classifier is a false negative (FN). If the classification method indicates that an observation is a \textit{Bot}, but the original label is \textit{Not}, then the classification is a false positive. Lastly, when the classification indicates that the observation is a \textit{Not}, whereas the original label indicates \textit{Bot}, it is deemed as a true negative. It is a oversimplification to only record the true positive rate of classifier accuracy. More nuanced measures of classification performance is therefore needed. There are many measures that can be derived from the confusion table, but only five measures will be calculated to assist in defining the most successful classification method among the 12 proposed. Table~\ref{T:Performance_Measures} briefly outlines the five accuracy measures. \footnote{The descriptions are from \citet{Sing2005}.}

\begin{table*}[]
	\centering
	\caption{Clustering Performance Measures Definition}
	\label{T:Performance_Measures}
	\begin{tabular}{@{}p{0.2\textwidth}p{0.15\textwidth}p{0.59\textwidth}@{}}
		\toprule
		\textbf{Measure}    & \textbf{Abbreviation} & \textbf{Description}  \\ \midrule
		False Positive Rate & FPR                  & The predictive outcome of the classifiers we used to identify if the instance is either bot or not. In this case the the prediction is positive, but the instance is negative. This is also known as fallout.                    \\
		True Positive Rate  & TPR                  & Similar to the FPR, however the prediction is correct. The prediction by the classifier and the instance were both positive. Calculation for this is done by dividing the true positives by the positive samples. This is also known as recall.                                   \\
		Accuracy            & Acc.                  & The degree to which the prediction made by the classifier matches the instance of the dataset, in this case Varol. To calculate the accuracy, the true positive and negative results are combined, and then divided by the positive and negative results.                          \\
		Phi Coefficient     & Phi                  & Supplies a range between minus one and one, where one is perfect prediction and zero is random prediction. Below zero is worse than random prediction.                    \\
		F-Score             & F                    & Introduced by van Rijsbergen in 1979, is a combination of precision and recall, and tests the accuracy of the test. The highest score for the F-Score is one and the worst score is zero.                                  \\
		Precision           & Prec.                 & Measures the probability that a classifier's prediction which has a positive outcome also has a positive instance. The reverse is true for a negative outcome.                      \\
		\bottomrule
	\end{tabular}
\end{table*}

\section{Results}
Each clustering method produced two clusters in which all observations are classified. It is not known which cluster is bot and which is not. The accuracy of each methodology is therefore first calculated to make an informed choice as to which cluster corresponds to which classification. In all the methodologies the `second' cluster was mostly grouping bots (as per the original labels), except four methods where the assumption is reversed: (1) Spearman-FANNY(full); (2) Spearman-PAM (full); (3) Pearson-FANNY (full); and (4) Pearson-PAM (full). The results reported below adjusted the assumptions appropriately.

The performance measures are captured in Table~\ref{T:Clust}. It is observed that the Spearman resulted in overall higher scores. On average, Spearman exhibits a slightly lower difference in FPR (0.04), which is the same score as Pearson for TPR. When considering phi, Spearman offers a marginal improvement over Pearson. The performance measurements for K-1 graphs performs between 0.10 and 0.23 points lower than that of the K-2 graph for Pearson, and between 0 and 0.19 lower for Spearman.

To further investigate the performance of the clustering methods, a receiver operating characteristics (ROC) graph is used to aid in the discussion. The ROC is commonly used graph in determining the general performance of clustering methods.

\begin{table}[]
	\begin{threeparttable}[c]
		\centering
		\footnotesize
		\caption{Clustering Performance Measures}
		\label{T:Clust}
		\setlength\tabcolsep{0.11cm}
		\begin{tabular}{@{}lllcccccc@{}}
			\toprule
			\multicolumn{1}{c}{\textbf{\begin{tabular}[c]{@{}c@{}}Distance \\ Method\end{tabular}}} & \multicolumn{1}{c}{\textbf{\begin{tabular}[c]{@{}c@{}}Graph \\ Type\end{tabular}}} & \multicolumn{1}{c}{\textbf{\begin{tabular}[c]{@{}c@{}}Clustering \\ Method\end{tabular}}} & \textbf{FPR}\tnote{a} & \textbf{TPR}\tnote{b} & \textbf{Acc}\tnote{c} & \textbf{Phi}\tnote{d} & \textbf{F}\tnote{e} & \textbf{Prec}\tnote{f} \\ \midrule
			\textbf{Spearman}\tnote{*} & \textbf{} & \textbf{} & \textbf{0.43} & \textbf{0.71} & \textbf{0.62} & \textbf{0.27} & \textbf{0.53} & \textbf{0.43} \\ \cmidrule(l){2-9}
			\textit{} & \textit{Original (K-2)}\tnote{*} & \textit{} & \textit{0.43} & \textit{0.82} & \textit{0.65} & \textit{0.37} & \textit{0.60} & \textit{0.47} \\ \cmidrule(l){2-9}
			&  & AGNES & 0.37 & 0.85 & 0.70 & 0.44 & 0.64 & 0.51 \\
			&  & FANNY & 0.42 & 0.77 & 0.64 & 0.32 & 0.57 & 0.46 \\
			&  & PAM & 0.49 & 0.85 & 0.62 & 0.35 & 0.58 & 0.44 \\ \cmidrule(l){2-9}
			\textit{} & \textit{K-1}\tnote{*} & \textit{} & \textit{0.43} & \textit{0.63} & \textit{0.59} & \textit{0.19} & \textit{0.49} & \textit{0.40} \\ \cmidrule(l){2-9}
			&  & AGNES & 0.30 & 0.46 & 0.62 & 0.16 & 0.44 & 0.41 \\
			&  & FANNY & 0.46 & 0.69 & 0.59 & 0.22 & 0.51 & 0.41 \\
			&  & PAM & 0.53 & 0.74 & 0.56 & 0.20 & 0.51 & 0.39 \\ \midrule
			\textbf{Pearson}\tnote{*} & \textbf{} & \textbf{} & \textbf{0.47} & \textbf{0.71} & \textbf{0.59} & \textbf{0.23} & \textbf{0.52} & \textbf{0.41} \\ \cmidrule(l){2-9}
			\textit{} & \textit{Original (K-2)}\tnote{*} & \textit{} & \textit{0.41} & \textit{0.79} & \textit{0.65} & \textit{0.36} & \textit{0.59} & \textit{0.47} \\ \cmidrule(l){2-9}
			&  & AGNES & 0.33 & 0.76 & 0.70 & 0.40 & 0.61 & 0.51 \\
			&  & FANNY & 0.42 & 0.77 & 0.64 & 0.32 & 0.57 & 0.46 \\
			&  & PAM & 0.49 & 0.85 & 0.62 & 0.35 & 0.58 & 0.44 \\ \cmidrule(l){2-9}
			\textit{} & \textit{K-1}\tnote{*} & \textit{} & \textit{0.51} & \textit{0.65} & \textit{0.54} & \textit{0.13} & \textit{0.47} & \textit{0.37} \\ \cmidrule(l){2-9}
			&  & AGNES & 0.54 & 0.51 & 0.48 & -0.03 & 0.38 & 0.30 \\
			&  & FANNY & 0.46 & 0.69 & 0.59 & 0.22 & 0.51 & 0.41 \\
			&  & PAM & 0.53 & 0.74 & 0.56 & 0.20 & 0.51 & 0.39 \\ \bottomrule
		\end{tabular}
		\begin{tablenotes}
			\item * Average of performance measures for the methods within the category.\\
			\item (a) False Positive Rate;
			\item (b) True Positive Rate;
			\item (c) Accuracy;
			\item (d) Phi Coefficient;
			\item (e) F-Score;
			\item (f) Precision;
		\end{tablenotes}
	\end{threeparttable}
\end{table}

\subsection{ROC}

\begin{figure}
	\includegraphics[width=0.48\textwidth]{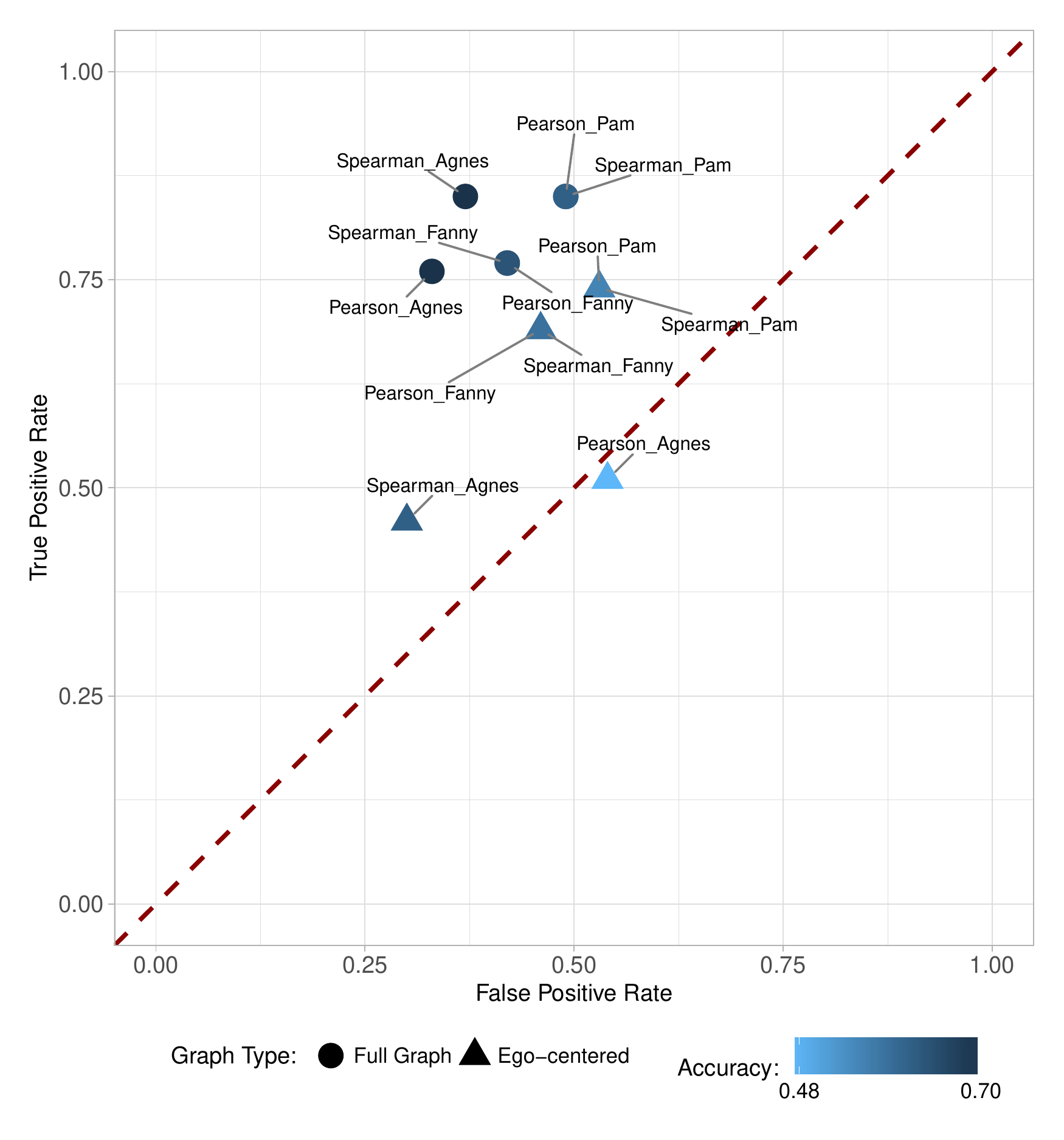}
	\caption{Receiver Operating Characteristics Plot}
	\label{F:ROC}
\end{figure}

The ROC graph visualises the classifiers based on their performance. The x and y scales display a range from zero to one. The false positive rate of the classifier is represented by x, and y represents the true positive rate. Coordinates $(0,1)$ or $(1,0)$ indicate the best performance, although the latter indicates an incorrect assumption of the classifier predictions compared to the original labels.\footnote{Since the assumption is adjusted, all methods would be above the x and y line.} The diagonal, $ \text{where } y = x$, indicated as a red dashed line on Figure~\ref{F:ROC}, displays a random guessing strategy. Any point on that line is considered to be equal to a random guessing strategy \citep{Fawcett2003}.

Figure~\ref{F:ROC} visualizes the clustering performance of the 12 methods. It is clear that the topology measures of the full graph were best able to contribute to true positive classification. The reduced graph negatively impacted the classification performance in all cases except one, Spearman-AGNES, which had a marginal improvement in false positive rates, but at high cost of true positives. Pearson-AGNES on the reduced graph is almost equal to a random guessing strategy, where the same method on the full graph contends for best classification methodology. Our hypothesis of a reduced K-1 graph leading to more accurate classification is, therefore, mostly rejected.

The remainder of the methods performed above the random guessing strategy. The methods with the best performance are: Spearman-AGNES; Pearson-PAM; and Spearman-PAM, all using the full K-2 graph topology.

The hypothesis that a K-2 network topology is capable of distinguishing between bots and humans is, therefore, confirmed.

\section{Conclusion}
Instead of using the network topology of an ego to classify the rest of the network as bot or not, as with Sibyl research, we use the network as the classifier of the ego itself. This relies on the intuition that bots are not capable of mimicking complex social network structures. While it is possible to classify bots using social network analysis with an estimated maximum accuracy of 0.70, it would best be used in conjunction with existing blended methodologies as a specialised method.

Of the five distance measures that were considered, Pearson and Spearman were chosen based on a visual inspection of their cluster definition. They offered the best definition of clusters and would offer the most accurate analysis of clusters. Four stability validation measures provided the three most applicable clustering methods: PAM, Fanny and AGNES. The combination of these distance and cluster analysis produced twelve methods used to build the classification model. Each of the methods were tested by calculating the difference between the predictions of each proposed method, and the labels provided by the Varol dataset. The labels were assigned to 1 based on the user having a classification score of 0.5 and above \citep{Varol2017}. In terms of correctly identifying Bots, three methods achieved an 85\% True Positive Rate. When also accounting for the False Positive Rate, two methods, achieved an accuracy score of 70\%. These two methods both employed the AGNES clustering method, on the full graph topography, and the choice of distance measure thus made no difference. Both these methods also performed exactly the same in terms of precision, with 51\%. The only measure to differentiate between the two measures is the Phi Coefficient. The Phi indicates that Spearman-AGNES, the highest performing measurement, performs 0.44 higher than that of random chance, which is impressive considering that the method is unsupervised.

Future methods should, therefore, consider including the social network topology, specifically a full K-2 graph of each user, as a means to successful classifying automated social agents on OSNs. We also note that we included all observations in the classification methodology. The methodology might be improved by removing obvious outliers, such as graphs containing only a few nodes. The inclusion of these spurious observations may affect the classification capabilities of the method.

\balance
\bibliographystyle{ACM-Reference-Format}
\bibliography{CSSG_SBT.bib}


\begin{thebibliography}{41}


\ifx \showCODEN    \undefined \def \showCODEN     #1{\unskip}     \fi
\ifx \showDOI      \undefined \def \showDOI       #1{#1}\fi
\ifx \showISBNx    \undefined \def \showISBNx     #1{\unskip}     \fi
\ifx \showISBNxiii \undefined \def \showISBNxiii  #1{\unskip}     \fi
\ifx \showISSN     \undefined \def \showISSN      #1{\unskip}     \fi
\ifx \showLCCN     \undefined \def \showLCCN      #1{\unskip}     \fi
\ifx \shownote     \undefined \def \shownote      #1{#1}          \fi
\ifx \showarticletitle \undefined \def \showarticletitle #1{#1}   \fi
\ifx \showURL      \undefined \def \showURL       {\relax}        \fi
\providecommand\bibfield[2]{#2}
\providecommand\bibinfo[2]{#2}
\providecommand\natexlab[1]{#1}
\providecommand\showeprint[2][]{arXiv:#2}

\bibitem[\protect\citeauthoryear{Aiello, Deplano, Schifanella, and
  Ruffo}{Aiello et~al\mbox{.}}{2012}]%
        {Aiello2014}
\bibfield{author}{\bibinfo{person}{Luca~Maria Aiello}, \bibinfo{person}{Martina
  Deplano}, \bibinfo{person}{Rossano Schifanella}, {and}
  \bibinfo{person}{Giancarlo Ruffo}.} \bibinfo{year}{2012}\natexlab{}.
\newblock \showarticletitle{{People are Strange when you're a Stranger: Impact
  and Influence of Bots on Social Networks}}. In
  \bibinfo{booktitle}{\emph{Proceedings of the Sixth International AAAI
  Conference on Weblogs and Social Media People}}. \bibinfo{address}{Dublin,
  Ireland}, \bibinfo{pages}{10--17}.
\newblock
\showISBNx{9781577355564}


\bibitem[\protect\citeauthoryear{Albert and Barab{\'{a}}si}{Albert and
  Barab{\'{a}}si}{2002}]%
        {Barabasi2002}
\bibfield{author}{\bibinfo{person}{Réka Albert} {and}
  \bibinfo{person}{Albert-László Barab{\'{a}}si}.}
  \bibinfo{year}{2002}\natexlab{}.
\newblock \showarticletitle{{Statistical mechanics of complex networks}}.
\newblock \bibinfo{journal}{\emph{Reviews of Modern Physics}}
  \bibinfo{volume}{74}, \bibinfo{number}{1} (\bibinfo{year}{2002}),
  \bibinfo{pages}{47--97}.
\newblock
\showISBNx{0034-6861}
\showISSN{0034-6861}
\urldef\tempurl%
\url{https://doi.org/10.1103/RevModPhys.74.47}
\showDOI{\tempurl}
\showeprint[arxiv]{cond-mat/0106096v1}


\bibitem[\protect\citeauthoryear{Bastos and Mercea}{Bastos and Mercea}{2017}]%
        {Bastos2017}
\bibfield{author}{\bibinfo{person}{Marco~T. Bastos} {and} \bibinfo{person}{Dan
  Mercea}.} \bibinfo{year}{2017}\natexlab{}.
\newblock \showarticletitle{{The Brexit Botnet and User-Generated Hyperpartisan
  News}}.
\newblock \bibinfo{journal}{\emph{Social Science Computer Review Preprint}}
  (\bibinfo{year}{2017}), \bibinfo{pages}{1--18}.
\newblock
\showISBNx{0954016157}
\showISSN{15528286}
\urldef\tempurl%
\url{https://doi.org/10.1177/0894439317734157}
\showDOI{\tempurl}


\bibitem[\protect\citeauthoryear{Bezdek and Hathaway}{Bezdek and
  Hathaway}{2002}]%
        {Bezdek2002}
\bibfield{author}{\bibinfo{person}{J.C. Bezdek} {and} \bibinfo{person}{R.J.
  Hathaway}.} \bibinfo{year}{2002}\natexlab{}.
\newblock \showarticletitle{{VAT: A Tool for Visual Assessment of (Cluster)
  Tendency}}. In \bibinfo{booktitle}{\emph{Proceedings of the 2002
  International Joint Conference on Neural Networks. IJCNN'02}}.
  \bibinfo{publisher}{IEEE}, \bibinfo{address}{Honolulu, HI},
  \bibinfo{pages}{2225--2230}.
\newblock
\showISBNx{0-7803-7278-6}
\showISSN{1098-7576}
\urldef\tempurl%
\url{https://doi.org/10.1109/IJCNN.2002.1007487}
\showDOI{\tempurl}


\bibitem[\protect\citeauthoryear{Bilge, Strufe, Balzarotti, and Kirda}{Bilge
  et~al\mbox{.}}{2009}]%
        {Bilge2009}
\bibfield{author}{\bibinfo{person}{Leyla Bilge}, \bibinfo{person}{Thorsten
  Strufe}, \bibinfo{person}{Davide Balzarotti}, {and} \bibinfo{person}{Engin
  Kirda}.} \bibinfo{year}{2009}\natexlab{}.
\newblock \showarticletitle{{All Your Contacts Are Belong to Us : Automated
  Identity Theft Attacks on Social Networks}}. In \bibinfo{booktitle}{\emph{WWW
  2009 MADRID!}} \bibinfo{address}{Mardrid}, \bibinfo{pages}{551--560}.
\newblock
\showISBNx{9781605584874}
\showISSN{15566013}
\urldef\tempurl%
\url{https://doi.org/10.1145/1526709.1526784}
\showDOI{\tempurl}


\bibitem[\protect\citeauthoryear{Boshmaf, Muslukhov, Beznosov, and
  Ripeanu}{Boshmaf et~al\mbox{.}}{2011}]%
        {Boshmaf2011}
\bibfield{author}{\bibinfo{person}{Yazan Boshmaf}, \bibinfo{person}{Ildar
  Muslukhov}, \bibinfo{person}{Konstantin Beznosov}, {and}
  \bibinfo{person}{Matei Ripeanu}.} \bibinfo{year}{2011}\natexlab{}.
\newblock \showarticletitle{{The socialbot network: when bots socialize for
  fame and money}}. In \bibinfo{booktitle}{\emph{Proceedings of the 27th annual
  computer security applications conference}}. \bibinfo{pages}{93--102}.
\newblock
\showISBNx{9781450306720}
\urldef\tempurl%
\url{https://doi.org/10.1145/2076732.2076746}
\showDOI{\tempurl}


\bibitem[\protect\citeauthoryear{Cresci, Di~Pietro, Petrocchi, Spognardi, and
  Tesconi}{Cresci et~al\mbox{.}}{2017}]%
        {Cresci2017}
\bibfield{author}{\bibinfo{person}{Stefano Cresci}, \bibinfo{person}{Roberto
  Di~Pietro}, \bibinfo{person}{Marinella Petrocchi}, \bibinfo{person}{Angelo
  Spognardi}, {and} \bibinfo{person}{Maurizio Tesconi}.}
  \bibinfo{year}{2017}\natexlab{}.
\newblock \showarticletitle{{Social Fingerprinting: detection of spambot groups
  through DNA-inspired behavioral modeling Stefano}}. In
  \bibinfo{booktitle}{\emph{Proceedings of the 26th International Conference on
  World Wide Web Companion}}. \bibinfo{address}{Perth, Australia},
  \bibinfo{pages}{963--972}.
\newblock
\showISBNx{9781450349147}
\showISSN{16130073}
\urldef\tempurl%
\url{https://doi.org/10.1145/3041021.3055135}
\showDOI{\tempurl}


\bibitem[\protect\citeauthoryear{Danezis and Mittal}{Danezis and
  Mittal}{2009}]%
        {Danezis2009}
\bibfield{author}{\bibinfo{person}{George Danezis} {and}
  \bibinfo{person}{Prateek Mittal}.} \bibinfo{year}{2009}\natexlab{}.
\newblock \showarticletitle{{SybilInfer: Detecting Sybil Nodes using Social
  Networks}}. In \bibinfo{booktitle}{\emph{16th Network {\&} Distributed System
  Security Symposium(NDSS)}}. \bibinfo{address}{San Diego, California, USA},
  \bibinfo{pages}{1--15}.
\newblock


\bibitem[\protect\citeauthoryear{Davis, Varol, Ferrara, Flammini, and
  Menczer}{Davis et~al\mbox{.}}{2016}]%
        {Davis2016}
\bibfield{author}{\bibinfo{person}{Clayton~Allen Davis}, \bibinfo{person}{Onur
  Varol}, \bibinfo{person}{Emilio Ferrara}, \bibinfo{person}{Alessandro
  Flammini}, {and} \bibinfo{person}{Filippo Menczer}.}
  \bibinfo{year}{2016}\natexlab{}.
\newblock \showarticletitle{{BotOrNot: A System to Evaluate Social Bots}}. In
  \bibinfo{booktitle}{\emph{Proceedings of the 25th International Conference
  Companion on World Wide Web}}. \bibinfo{publisher}{ACM Press},
  \bibinfo{address}{Montr{\'{e}}al, Qu{\'{e}}bec, Canada},
  \bibinfo{pages}{273--274}.
\newblock
\showISBNx{10638016}
\showISSN{10638016}
\urldef\tempurl%
\url{https://doi.org/10.1145/2872518.2889302}
\showDOI{\tempurl}


\bibitem[\protect\citeauthoryear{Douceur}{Douceur}{2002}]%
        {Douceur2002}
\bibfield{author}{\bibinfo{person}{John~R Douceur}.}
  \bibinfo{year}{2002}\natexlab{}.
\newblock \showarticletitle{{The Sybil Attack}}. In
  \bibinfo{booktitle}{\emph{IPTPS '01 Revised Papers from the First
  International Workshop on Peer-to-Peer Systems}}.
  \bibinfo{publisher}{Springer}, \bibinfo{address}{London, England},
  \bibinfo{pages}{251--260}.
\newblock
\showISBNx{3540441794}
\showISSN{00278424}


\bibitem[\protect\citeauthoryear{Echeverr{\'{i}}a and Zhou}{Echeverr{\'{i}}a
  and Zhou}{2017}]%
        {Echeverria2017}
\bibfield{author}{\bibinfo{person}{Juan Echeverr{\'{i}}a} {and}
  \bibinfo{person}{Shi Zhou}.} \bibinfo{year}{2017}\natexlab{}.
\newblock \showarticletitle{{Discovery, Retrieval, and Analysis of 'Star Wars'
  botnet in Twitter}}. In \bibinfo{booktitle}{\emph{Proceedings of the 2017
  IEEE/ACM International Conference on Advances in Social Networks Analysis and
  Mining 2017}}. \bibinfo{address}{Sydney, Australia}, \bibinfo{pages}{1--8}.
\newblock
\showISBNx{9781450349932}
\urldef\tempurl%
\url{https://doi.org/10.1145/3110025.3110074}
\showDOI{\tempurl}


\bibitem[\protect\citeauthoryear{Edwards, Edwards, Spence, and Shelton}{Edwards
  et~al\mbox{.}}{2014}]%
        {Edwards2014}
\bibfield{author}{\bibinfo{person}{Chad Edwards}, \bibinfo{person}{Autumn
  Edwards}, \bibinfo{person}{Patric~R. Spence}, {and}
  \bibinfo{person}{Ashleigh~K. Shelton}.} \bibinfo{year}{2014}\natexlab{}.
\newblock \showarticletitle{{Is that a bot running the social media feed?
  Testing the differences in perceptions of communication quality for a human
  agent and a bot agent on Twitter}}.
\newblock \bibinfo{journal}{\emph{Computers in Human Behavior}}
  \bibinfo{volume}{33} (\bibinfo{year}{2014}), \bibinfo{pages}{372--376}.
\newblock
\showISBNx{0747-5632}
\showISSN{07475632}
\urldef\tempurl%
\url{https://doi.org/10.1016/j.chb.2013.08.013}
\showDOI{\tempurl}


\bibitem[\protect\citeauthoryear{Fawcett}{Fawcett}{2003}]%
        {Fawcett2003}
\bibfield{author}{\bibinfo{person}{Tom Fawcett}.}
  \bibinfo{year}{2003}\natexlab{}.
\newblock \bibinfo{booktitle}{\emph{{ROC Graphs: Notes and Practical
  Considerations for Data Mining Researchers ROC Graphs : Notes and Practical
  Considerations for Data Mining Researchers}}}.
\newblock \bibinfo{type}{{T}echnical {R}eport}.
  \bibinfo{institution}{Intelligent Enterprise Technologies Laboratory},
  \bibinfo{address}{Palo Alto}. \bibinfo{pages}{27} pages.
\newblock
\showISBNx{Technical report}
\showISSN{08997667}
\urldef\tempurl%
\url{https://doi.org/10.1.1.10.9777}
\showDOI{\tempurl}


\bibitem[\protect\citeauthoryear{Ferguson}{Ferguson}{2017}]%
        {Ferguson2017}
\bibfield{author}{\bibinfo{person}{Niall Ferguson}.}
  \bibinfo{year}{2017}\natexlab{}.
\newblock \showarticletitle{{The False Prophecy of Hyperconnection: How to
  Survive the Networked Age}}.
\newblock \bibinfo{journal}{\emph{Foreign Affairs}} \bibinfo{volume}{96},
  \bibinfo{number}{68} (\bibinfo{year}{2017}), \bibinfo{pages}{8--23}.
\newblock
\showISBNx{1120617163123}
\showISSN{02729490}
\urldef\tempurl%
\url{https://doi.org/10.3868/s050-004-015-0003-8}
\showDOI{\tempurl}


\bibitem[\protect\citeauthoryear{Ferrara, Varol, Davis, Menczer, and
  Flammini}{Ferrara et~al\mbox{.}}{2016}]%
        {Ferrara2016}
\bibfield{author}{\bibinfo{person}{Emilio Ferrara}, \bibinfo{person}{Onur
  Varol}, \bibinfo{person}{Clayton Davis}, \bibinfo{person}{Filippo Menczer},
  {and} \bibinfo{person}{Alessandro Flammini}.}
  \bibinfo{year}{2016}\natexlab{}.
\newblock \showarticletitle{{The Rise of Social Bots}}.
\newblock \bibinfo{journal}{\emph{Commun. ACM}} \bibinfo{volume}{59},
  \bibinfo{number}{7} (\bibinfo{date}{6} \bibinfo{year}{2016}),
  \bibinfo{pages}{96--104}.
\newblock
\showISSN{00010782}
\urldef\tempurl%
\url{https://doi.org/10.1145/2818717}
\showDOI{\tempurl}


\bibitem[\protect\citeauthoryear{Findlay}{Findlay}{2018}]%
        {Findlay2018}
\bibfield{author}{\bibinfo{person}{Kyle Findlay}.}
  \bibinfo{year}{2018}\natexlab{}.
\newblock \bibinfo{title}{{South Africans take Bell Pottinger's interference
  personally: reactions to the PR company on Twitter}}.
\newblock , \bibinfo{numpages}{11}~pages.
\newblock
\urldef\tempurl%
\url{https://bit.ly/2MFbTNa}
\showURL{%
\tempurl}


\bibitem[\protect\citeauthoryear{Fraser}{Fraser}{2017}]%
        {Fraser2017}
\bibfield{author}{\bibinfo{person}{Andrew Fraser}.}
  \bibinfo{year}{2017}\natexlab{}.
\newblock \bibinfo{title}{{TechCentral: We go inside the Guptabot fake news
  network}}.
\newblock , \bibinfo{numpages}{10}~pages.
\newblock
\urldef\tempurl%
\url{https://bit.ly/2NxMRvK}
\showURL{%
\tempurl}


\bibitem[\protect\citeauthoryear{Gao, Wang, Gong, Kulkrani, Thomas, and
  Mittal}{Gao et~al\mbox{.}}{2018}]%
        {Gao2018}
\bibfield{author}{\bibinfo{person}{Peng Gao}, \bibinfo{person}{Binghui Wang},
  \bibinfo{person}{Neil~Zhenqiang Gong}, \bibinfo{person}{Sanjeev~R Kulkrani},
  \bibinfo{person}{Kurt Thomas}, {and} \bibinfo{person}{Prateek Mittal}.}
  \bibinfo{year}{2018}\natexlab{}.
\newblock \bibinfo{title}{{SybilFuse : Combining Local Attributes with Global
  Structure to Perform Robust Sybil Detection}}.  (\bibinfo{year}{2018}).
\newblock
\showISBNx{1234567245}


\bibitem[\protect\citeauthoryear{Gerbaudo}{Gerbaudo}{2012}]%
        {Gerbaudo2012}
\bibfield{author}{\bibinfo{person}{Paolo Gerbaudo}.}
  \bibinfo{year}{2012}\natexlab{}.
\newblock \bibinfo{booktitle}{\emph{{Tweets and the Streets: Social Media and
  Contemporary Activism}}}.
\newblock \bibinfo{publisher}{Pluto Press}, \bibinfo{address}{London, England}.
  196 pages.
\newblock
\showISBNx{978 1 8496 4800 4}
\showISSN{1098-6596}
\urldef\tempurl%
\url{https://doi.org/10.1017/CBO9781107415324.004}
\showDOI{\tempurl}


\bibitem[\protect\citeauthoryear{Ghosh, Viswanath, Kooti, Sharma, Korlam,
  Benevenuto, Ganguly, and Gummadi}{Ghosh et~al\mbox{.}}{2012}]%
        {Ghosh2012}
\bibfield{author}{\bibinfo{person}{Saptarshi Ghosh}, \bibinfo{person}{Bimal
  Viswanath}, \bibinfo{person}{Farshad Kooti}, \bibinfo{person}{Naveen~Kumar
  Sharma}, \bibinfo{person}{Gautam Korlam}, \bibinfo{person}{Fabricio
  Benevenuto}, \bibinfo{person}{Niloy Ganguly}, {and}
  \bibinfo{person}{Krishna~Phani Gummadi}.} \bibinfo{year}{2012}\natexlab{}.
\newblock \showarticletitle{{Understanding and combating link farming in the
  twitter social network}}. In \bibinfo{booktitle}{\emph{Proceedings of the
  21st international conference on World Wide Web - WWW '12}}.
  \bibinfo{address}{Lyon,France}, \bibinfo{pages}{61--70}.
\newblock
\showISBNx{9781450312295}
\urldef\tempurl%
\url{https://doi.org/10.1145/2187836.2187846}
\showDOI{\tempurl}


\bibitem[\protect\citeauthoryear{Gilani, Kochmar, and Crowcroft}{Gilani
  et~al\mbox{.}}{2017}]%
        {Gilani2017}
\bibfield{author}{\bibinfo{person}{Zafar Gilani}, \bibinfo{person}{Ekaterina
  Kochmar}, {and} \bibinfo{person}{Jon Crowcroft}.}
  \bibinfo{year}{2017}\natexlab{}.
\newblock \showarticletitle{{Classification of Twitter Accounts into Automated
  Agents and Human Users}}. In \bibinfo{booktitle}{\emph{Proceedings of the
  2017 IEEE/ACM International Conference on Advances in Social Networks
  Analysis and Mining 2017}}. \bibinfo{publisher}{ACM},
  \bibinfo{address}{Sydney, Australia}, \bibinfo{pages}{489--496}.
\newblock
\showISBNx{9781450349932}
\urldef\tempurl%
\url{https://doi.org/10.1145/3110025.3110091}
\showDOI{\tempurl}


\bibitem[\protect\citeauthoryear{Harris, Hirst, and Mossinghoff}{Harris
  et~al\mbox{.}}{2008}]%
        {Harris2008}
\bibfield{author}{\bibinfo{person}{John Harris}, \bibinfo{person}{Jeffry~L.
  Hirst}, {and} \bibinfo{person}{Michael Mossinghoff}.}
  \bibinfo{year}{2008}\natexlab{}.
\newblock \showarticletitle{{Combinatorics and Graph Theory}}.
\newblock In \bibinfo{booktitle}{\emph{Combinatorics and graph theory (Vol. 2,
  p. 139)} (\bibinfo{edition}{2} ed.)}. \bibinfo{publisher}{Springer},
  \bibinfo{address}{New York, NY}, Chapter 1: Graph T, \bibinfo{pages}{1--30}.
\newblock
\showISBNx{978-0-387-79710-6}
\showISSN{0172-6056}
\urldef\tempurl%
\url{https://doi.org/10.1007/978-0-387-79711-3}
\showDOI{\tempurl}


\bibitem[\protect\citeauthoryear{Howard, Duffy, Freelon, Hussain, Mari, and
  Mazaid}{Howard et~al\mbox{.}}{2011}]%
        {Howard2011}
\bibfield{author}{\bibinfo{person}{Philip~N Howard}, \bibinfo{person}{Aiden
  Duffy}, \bibinfo{person}{Deen Freelon}, \bibinfo{person}{Muzammil Hussain},
  \bibinfo{person}{Will Mari}, {and} \bibinfo{person}{Mrawa Mazaid}.}
  \bibinfo{year}{2011}\natexlab{}.
\newblock \bibinfo{booktitle}{\emph{{Opening closed regimes: what was the role
  of social media during the Arab Spring?}}}
\newblock \bibinfo{type}{{T}echnical {R}eport}. \bibinfo{pages}{1--30} pages.
\newblock
\showISBNx{9780874216561}
\showISSN{0717-6163}
\urldef\tempurl%
\url{https://doi.org/10.1007/s13398-014-0173-7.2}
\showDOI{\tempurl}
\showeprint[arxiv]{gr-qc/9809069v1}


\bibitem[\protect\citeauthoryear{Italiano, Laura, and Santaroni}{Italiano
  et~al\mbox{.}}{2012}]%
        {Italiano2012}
\bibfield{author}{\bibinfo{person}{Giuseppe~F. Italiano},
  \bibinfo{person}{Luigi Laura}, {and} \bibinfo{person}{Federico Santaroni}.}
  \bibinfo{year}{2012}\natexlab{}.
\newblock \showarticletitle{{Finding strong bridges and strong articulation
  points in linear time}}.
\newblock \bibinfo{journal}{\emph{Theoretical Computer Science}}
  \bibinfo{volume}{447} (\bibinfo{year}{2012}), \bibinfo{pages}{74--84}.
\newblock
\showISBNx{3642174574}
\showISSN{03043975}
\urldef\tempurl%
\url{https://doi.org/10.1016/j.tcs.2011.11.011}
\showDOI{\tempurl}


\bibitem[\protect\citeauthoryear{John}{John}{2018}]%
        {Gramlich2018}
\bibfield{author}{\bibinfo{person}{Gramlich John}.}
  \bibinfo{year}{2018}\natexlab{}.
\newblock \bibinfo{title}{{Q And A: How Pew Research Center identified bots on
  Twitter}}.
\newblock
\newblock
\urldef\tempurl%
\url{http://www.pewresearch.org/fact-tank/2018/04/19/qa-how-pew-research-center-identified-bots-on-twitter/}
\showURL{%
\tempurl}


\bibitem[\protect\citeauthoryear{Kassambara}{Kassambara}{2017}]%
        {Kassambara2017}
\bibfield{author}{\bibinfo{person}{Alboukadel Kassambara}.}
  \bibinfo{year}{2017}\natexlab{}.
\newblock \bibinfo{booktitle}{\emph{{Practical Guide to Cluster Analysis in R:
  Unsupervised Machine Learning}} (\bibinfo{edition}{1} ed.)}.
\newblock \bibinfo{publisher}{STHDA}. 187 pages.
\newblock
\showISBNx{1542462703}


\bibitem[\protect\citeauthoryear{Kwak, Lee, Park, and Moon}{Kwak
  et~al\mbox{.}}{2010}]%
        {Kwak2010}
\bibfield{author}{\bibinfo{person}{Haewoon Kwak}, \bibinfo{person}{Changhyun
  Lee}, \bibinfo{person}{Hosung Park}, {and} \bibinfo{person}{Sue Moon}.}
  \bibinfo{year}{2010}\natexlab{}.
\newblock \showarticletitle{{What is Twitter, a Social Network or a News Media
  ?}}. In \bibinfo{booktitle}{\emph{Proceedings of the 19th international
  conference on World Wide Web (IW3C2)}}. \bibinfo{address}{Raleigh, North
  Carolina, USA.}, \bibinfo{pages}{591--600}.
\newblock
\showISBNx{9781605587998}
\showISSN{1932-8036}
\urldef\tempurl%
\url{https://doi.org/10.1145/1772690.1772751}
\showDOI{\tempurl}


\bibitem[\protect\citeauthoryear{Newman}{Newman}{2003}]%
        {Newman2003}
\bibfield{author}{\bibinfo{person}{Mark E~J Newman}.}
  \bibinfo{year}{2003}\natexlab{}.
\newblock \showarticletitle{{Mixing patterns in networks}}.
\newblock \bibinfo{journal}{\emph{Physical Review E - Statistical Physics,
  Plasmas, Fluids, and Related Interdisciplinary Topics}} \bibinfo{volume}{67},
  \bibinfo{number}{2} (\bibinfo{year}{2003}), \bibinfo{pages}{13}.
\newblock
\showISBNx{1539-3755}
\showISSN{1063651X}
\urldef\tempurl%
\url{https://doi.org/10.1103/PhysRevE.67.026126}
\showDOI{\tempurl}


\bibitem[\protect\citeauthoryear{Newman}{Newman}{2010}]%
        {Newman2010}
\bibfield{author}{\bibinfo{person}{Mark E~J Newman}.}
  \bibinfo{year}{2010}\natexlab{}.
\newblock \bibinfo{booktitle}{\emph{{Networks: An Introduction}}}.
\newblock \bibinfo{publisher}{Oxford University Press}, \bibinfo{address}{New
  York, NY}. 1--784 pages.
\newblock
\showISBNx{9780191594175}
\showISSN{1578-1275}
\urldef\tempurl%
\url{https://doi.org/10.1093/acprof:oso/9780199206650.001.0001}
\showDOI{\tempurl}


\bibitem[\protect\citeauthoryear{Shao, Hui, Wang, Jiang, Flammini, Menczer, and
  Ciampaglia}{Shao et~al\mbox{.}}{2018}]%
        {Shao2018}
\bibfield{author}{\bibinfo{person}{Chengcheng Shao}, \bibinfo{person}{Pik-Mai
  Hui}, \bibinfo{person}{Lei Wang}, \bibinfo{person}{Xinwen Jiang},
  \bibinfo{person}{Alessandro Flammini}, \bibinfo{person}{Filippo Menczer},
  {and} \bibinfo{person}{Giovanni~Luca Ciampaglia}.}
  \bibinfo{year}{2018}\natexlab{}.
\newblock \showarticletitle{{Anatomy of an online misinformation network}}.
\newblock \bibinfo{journal}{\emph{PLoS ONE}} \bibinfo{volume}{13},
  \bibinfo{number}{4} (\bibinfo{year}{2018}), \bibinfo{pages}{1--23}.
\newblock
\showISBNx{1111111111}
\urldef\tempurl%
\url{https://doi.org/10.1371/journal.pone.0196087}
\showDOI{\tempurl}


\bibitem[\protect\citeauthoryear{Sing, Sander, Beerenwinkel, and Lengauer}{Sing
  et~al\mbox{.}}{2005}]%
        {Sing2005}
\bibfield{author}{\bibinfo{person}{Tobias Sing}, \bibinfo{person}{Oliver
  Sander}, \bibinfo{person}{Niko Beerenwinkel}, {and} \bibinfo{person}{Thomas
  Lengauer}.} \bibinfo{year}{2005}\natexlab{}.
\newblock \showarticletitle{{ROCR: Visualizing classifier performance in R}}.
\newblock \bibinfo{journal}{\emph{Bioinformatics}} \bibinfo{volume}{21},
  \bibinfo{number}{20} (\bibinfo{year}{2005}), \bibinfo{pages}{3940--3941}.
\newblock
\showISBNx{1367-4803 (Print){\textbackslash}n1367-4803 (Linking)}
\showISSN{13674803}
\urldef\tempurl%
\url{https://doi.org/10.1093/bioinformatics/bti623}
\showDOI{\tempurl}


\bibitem[\protect\citeauthoryear{Subrahmanian, Azaria, Durst, Kagan, Galstyan,
  Lerman, Zhu, Ferrara, Flammini, and Menczer}{Subrahmanian
  et~al\mbox{.}}{2016}]%
        {Subrahmanian2016}
\bibfield{author}{\bibinfo{person}{V.S. Subrahmanian}, \bibinfo{person}{Amos
  Azaria}, \bibinfo{person}{Skylar Durst}, \bibinfo{person}{Vadim Kagan},
  \bibinfo{person}{Aram Galstyan}, \bibinfo{person}{Kristina Lerman},
  \bibinfo{person}{Linhong Zhu}, \bibinfo{person}{Emilio Ferrara},
  \bibinfo{person}{Alessandro Flammini}, {and} \bibinfo{person}{Filippo
  Menczer}.} \bibinfo{year}{2016}\natexlab{}.
\newblock \showarticletitle{{The DARPA Twitter Bot Challenge}}.
\newblock \bibinfo{journal}{\emph{Computer}} \bibinfo{volume}{49},
  \bibinfo{number}{6} (\bibinfo{date}{6} \bibinfo{year}{2016}),
  \bibinfo{pages}{38--46}.
\newblock
\showISSN{0018-9162}
\urldef\tempurl%
\url{https://doi.org/10.1109/MC.2016.183}
\showDOI{\tempurl}


\bibitem[\protect\citeauthoryear{Varol, Ferrara, Davis, Menczer, and
  Flammini}{Varol et~al\mbox{.}}{2017}]%
        {Varol2017}
\bibfield{author}{\bibinfo{person}{Onur Varol}, \bibinfo{person}{Emilio
  Ferrara}, \bibinfo{person}{Clayton~A. Davis}, \bibinfo{person}{Filippo
  Menczer}, {and} \bibinfo{person}{Alessandro Flammini}.}
  \bibinfo{year}{2017}\natexlab{}.
\newblock \showarticletitle{{Online Human-Bot Interactions: Detection,
  Estimation, and Characterization}}. In \bibinfo{booktitle}{\emph{Proceedings
  of the Eleventh International AAAI Conference on Web and Social Media (ICWSM
  2017)}}. \bibinfo{publisher}{Association for the Advancement of Artificial
  Intelligence}, \bibinfo{address}{Montreal, Quebec, Canada},
  \bibinfo{pages}{280--289}.
\newblock
\showISBNx{9781577357889}
\urldef\tempurl%
\url{http://arxiv.org/abs/1703.03107}
\showURL{%
\tempurl}


\bibitem[\protect\citeauthoryear{Viswanath, Post, Gummadi, and
  Mislove}{Viswanath et~al\mbox{.}}{2010}]%
        {Viswanath2010}
\bibfield{author}{\bibinfo{person}{Bimal Viswanath}, \bibinfo{person}{Ansley
  Post}, \bibinfo{person}{Krishna~P. Gummadi}, {and} \bibinfo{person}{Alan
  Mislove}.} \bibinfo{year}{2010}\natexlab{}.
\newblock \showarticletitle{{An analysis of social network-based Sybil
  defenses}}. In \bibinfo{booktitle}{\emph{ACM SIGCOMM Computer Communication
  Review 2010}}. \bibinfo{address}{New Delhi, India},
  \bibinfo{pages}{363--374}.
\newblock
\showISBNx{9781450302012}
\showISSN{01464833}
\urldef\tempurl%
\url{https://doi.org/10.1145/1851182.1851226}
\showDOI{\tempurl}


\bibitem[\protect\citeauthoryear{Wagstaff, Cardie, Rogers, and
  Schroedl}{Wagstaff et~al\mbox{.}}{2001}]%
        {Wagstaff2001}
\bibfield{author}{\bibinfo{person}{Kiri Wagstaff}, \bibinfo{person}{Claire
  Cardie}, \bibinfo{person}{Seth Rogers}, {and} \bibinfo{person}{Stefan
  Schroedl}.} \bibinfo{year}{2001}\natexlab{}.
\newblock \showarticletitle{{Constrained K-means Clustering with Background
  Knowledge}}. In \bibinfo{booktitle}{\emph{Proceedings of the Eighteenth
  International Conference on Machine Learning}}.
  \bibinfo{address}{Williamstown, MA, USA}, \bibinfo{pages}{577--584}.
\newblock
\showISBNx{1558607781}
\showISSN{0162-8828}
\urldef\tempurl%
\url{https://doi.org/10.1109/TPAMI.2002.1017616}
\showDOI{\tempurl}


\bibitem[\protect\citeauthoryear{Wang, Mohanlal, Wilson, Wang, Metzger, Zheng,
  and Zhao}{Wang et~al\mbox{.}}{2012}]%
        {Wang2012}
\bibfield{author}{\bibinfo{person}{Gang Wang}, \bibinfo{person}{Manish
  Mohanlal}, \bibinfo{person}{Christo Wilson}, \bibinfo{person}{Xiao Wang},
  \bibinfo{person}{Miriam Metzger}, \bibinfo{person}{Haitao Zheng}, {and}
  \bibinfo{person}{Ben~Y. Zhao}.} \bibinfo{year}{2012}\natexlab{}.
\newblock \showarticletitle{{Social Turing Tests: Crowdsourcing Sybil
  Detection}}. In \bibinfo{booktitle}{\emph{arXiv preprint}}.
\newblock
\urldef\tempurl%
\url{http://arxiv.org/abs/1205.3856}
\showURL{%
\tempurl}


\bibitem[\protect\citeauthoryear{Wasserman and Faust}{Wasserman and
  Faust}{1994}]%
        {Wasserman1994}
\bibfield{author}{\bibinfo{person}{Stanley Wasserman} {and}
  \bibinfo{person}{Katherine Faust}.} \bibinfo{year}{1994}\natexlab{}.
\newblock \bibinfo{booktitle}{\emph{{Social Network Analysis: Methods and
  Applications}} (\bibinfo{edition}{first} ed.)}.
\newblock \bibinfo{publisher}{Cambridge University Press},
  \bibinfo{address}{New York}.
\newblock


\bibitem[\protect\citeauthoryear{Yang, Harkreader, and Gu}{Yang
  et~al\mbox{.}}{2013}]%
        {Yang2013}
\bibfield{author}{\bibinfo{person}{Chao Yang}, \bibinfo{person}{Robert
  Harkreader}, {and} \bibinfo{person}{Guofei Gu}.}
  \bibinfo{year}{2013}\natexlab{}.
\newblock \showarticletitle{{Empirical Evaluation and New Design for Fighting
  Evolving Twitter Spammers}}.
\newblock \bibinfo{journal}{\emph{IEEE Transactions on Information Forensics
  and Security}} \bibinfo{volume}{8}, \bibinfo{number}{8}
  (\bibinfo{year}{2013}), \bibinfo{pages}{1280–1293}.
\newblock
\urldef\tempurl%
\url{http://citeseerx.ist.psu.edu/viewdoc/download?doi=10.1.1.414.5888&rep=rep1&type=pdf}
\showURL{%
\tempurl}


\bibitem[\protect\citeauthoryear{Yang, Harkreader, Zhang, Shin, and Gu}{Yang
  et~al\mbox{.}}{2012}]%
        {Yang2012}
\bibfield{author}{\bibinfo{person}{Chao Yang}, \bibinfo{person}{Robert
  Harkreader}, \bibinfo{person}{Jialong Zhang}, \bibinfo{person}{Seungwon
  Shin}, {and} \bibinfo{person}{Guofei Gu}.} \bibinfo{year}{2012}\natexlab{}.
\newblock \showarticletitle{{Analyzing spammers' social networks for fun and
  profit: a case study of cyber criminal ecosystem on twitter}}. In
  \bibinfo{booktitle}{\emph{WWW '12: Proceedings of the 21st international
  conference on World Wide Web}}. \bibinfo{address}{Lyon,France},
  \bibinfo{pages}{71--80}.
\newblock
\showISBNx{9781450312295}
\urldef\tempurl%
\url{https://doi.org/10.1145/2187836.2187847}
\showDOI{\tempurl}


\bibitem[\protect\citeauthoryear{Yu, Gibbons, Kaminsky, and Xiao}{Yu
  et~al\mbox{.}}{2008}]%
        {Yu2008a}
\bibfield{author}{\bibinfo{person}{Haifeng Yu}, \bibinfo{person}{Phillip~B
  Gibbons}, \bibinfo{person}{Michael Kaminsky}, {and} \bibinfo{person}{Feng
  Xiao}.} \bibinfo{year}{2008}\natexlab{}.
\newblock \showarticletitle{{SybilLimit: A Near-Optimal Social Network Defense
  against Sybil Attacks}}. In \bibinfo{booktitle}{\emph{IEEE Symposium on
  Security and Privacy}}. \bibinfo{pages}{3--17}.
\newblock
\showISBNx{9780769531687}
\urldef\tempurl%
\url{https://doi.org/10.1109/SP.2008.13}
\showDOI{\tempurl}


\bibitem[\protect\citeauthoryear{Yu, Kaminsky, Gibbons, and Flaxman}{Yu
  et~al\mbox{.}}{2006}]%
        {Yu2008}
\bibfield{author}{\bibinfo{person}{Haifeng Yu}, \bibinfo{person}{Michael
  Kaminsky}, \bibinfo{person}{Philip~B. Gibbons}, {and}
  \bibinfo{person}{Abraham~D. Flaxman}.} \bibinfo{year}{2006}\natexlab{}.
\newblock \showarticletitle{{SybilGuard: Defending against sybil attacks via
  social networks}}.
\newblock \bibinfo{journal}{\emph{ACM SIGCOMM Computer Communication Review}}
  \bibinfo{volume}{36}, \bibinfo{number}{4} (\bibinfo{year}{2006}),
  \bibinfo{pages}{267--278}.
\newblock
\showISBNx{1595933085}
\showISSN{10636692}
\urldef\tempurl%
\url{https://doi.org/10.1109/TNET.2008.923723}
\showDOI{\tempurl}


\end{thebibliography}

\end{document}